\documentclass[runningheads]{llncs}
\usepackage[T1]{fontenc}
\usepackage{xcolor}
\usepackage{lipsum}
\usepackage{amsfonts}
\usepackage{algorithmic}
\usepackage{siunitx}
\usepackage{textcomp}
\usepackage{rotating,multirow}
\usepackage{color,soul}
\usepackage{graphicx}
\usepackage{threeparttable}
\usepackage{tabularx}
\usepackage{booktabs}
\usepackage{color, colortbl}

\usepackage{array}
\usepackage{wrapfig}
\begin{document}
\title{Mapping code \\on Coarse Grained Reconfigurable Arrays \\using a SAT solver}
%
%
\author{Cristian
Tirelli\inst{1}\orcidID{0009-0002-5403-6255} \and
Laura
Pozzi\inst{1}\orcidID{0000-0003-1083-8782} 
}
\authorrunning{F. Author et al.}
%
\institute{Università della Svizzera Italiana, Switzerland 
\email{\{tirelc,pozzil\}@usi.ch}}
\maketitle              
\begin{abstract}

Emerging low-powered architectures like Coarse-Grain Reconfigurable Arrays (CGRAs) are becoming more common. Often included as co-processors, they are used to accelerate compute-intensive workloads like loops. The speedup obtained is defined by the hardware design of the accelerator and by the quality of the compilation.
State of the art (SoA) compilation techniques leverage modulo scheduling to minimize the Iteration Interval (II), exploit the architecture parallelism and, consequentially, reduce the execution time of the accelerated workload.
In our work, we focus on improving the compilation process by finding the lowest II for any given topology, through a satisfiability (SAT) formulation of the mapping problem.
We introduce a novel schedule, called Kernel Mobility Schedule, to encode all the possible mappings for a given Data Flow Graph (DFG) and for a given II. The schedule is used together with the CGRA architectural information to generate all the constraints necessary to find a valid mapping. 
Experimental results demonstrate that our method not only reduces compilation time on average but also achieves higher quality mappings compared to existing SoA techniques.


\end{abstract}
\section{Problem Description}
\subsection{Introduction}
Achieving high efficiency with low power requirements is a recurring problem in hardware design and there are many architectures in the state-of-the-art trying to address this problem. A promising set of architectures that fits the specific are Coarse-Grain Reconfigurable Arrays (CGRAs)  \cite{li2021chordmap,karunaratne2017hycube}. 
They are often used as co-processor to accelerate the most intensive part of a code, e.g. loops, and their most common structure is depicted in Figure \ref{fig:cgra-map}.a.
Typically, every Processing Element (PE) is connected with the nearest PE and is equipped with an Arithmetic-Logic Unit (ALU) and some internal registers. The PE network allows the exchange of data produced during the computation by the processing elements, while the connection to the main memory allows every PE to load external data.
Good code generation is necessary to enable high performance on this kind of architecture. \\
The key challenge in the compilation process is to find a space-time mapping that can fully exploit the intrinsic instruction-level parallelism offered by the architecture.
Our research is currently focused on exploring new mapping techniques using a SAT-based approach \cite{Tirelli2023,sat_extended}, where constraints related to data dependency, architectural features, and scheduling are formulated as Boolean equations.

\subsection{Mapping problem}
Previous mapping techniques approached the problem in a different way. Some methods reduced the mapping problem to a graph theory problem \cite{dave2018ramp,chen2014graph},  while others solved the problem with linear programming \cite{chin2018architecture}. However, many of these methods produced mappings of low quality, or alternatively, high-quality mappings but with significantly extended compilation times.

\begin{figure}
    \centering
    \includegraphics[width=\textwidth]{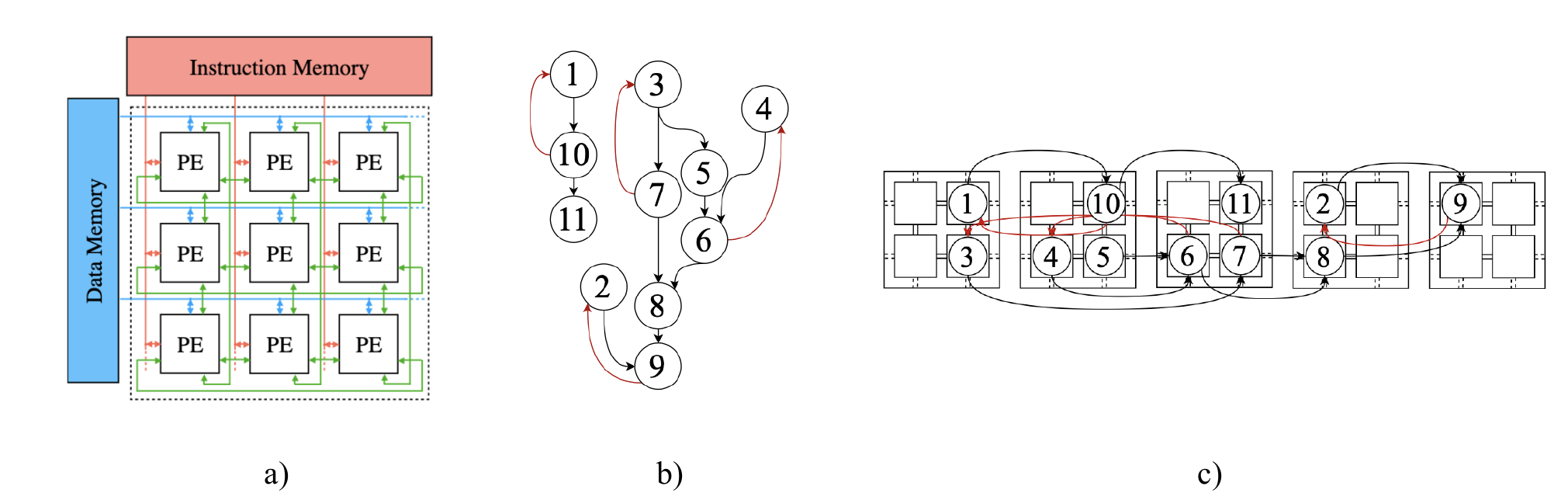}
    \caption{a) Abstract $3\times 3$ CGRA architecture with 2d-mesh topology. b) Loop in DFG form. Red edges are loop-carried dependencies; black edges are data dependencies. c) One valid mapping of the DFG on the left in a $2\times 2$ CGRA}
    \label{fig:cgra-map}
\end{figure}

The CGRA mapping problem requires finding a valid space-time mapping from a DFG to the target CGRA architecture by entailing different challenges. 
In the context of CGRA this implies assigning the program instructions to the device PE and routing the data guaranteeing the functional correctness of the application. Figure \ref{fig:cgra-map}.b depict a DFG and Figure \ref{fig:cgra-map}.c shows a possible mapping of it on a $2\times 2$ CGRA, without applying modulo scheduling \cite{rau1996iterative}.

\vskip -1em

\subsection{Modulo Scheduling}

\textit{Modulo scheduling} is a common compilation technique used to execute multiple iterations of a loop in an interleaved manner to reduce the execution time. 
After this optimization is applied, the code mapped on the device is divided into three sections: prologue, kernel and epilogue. As shown in Figure \ref{fig:kms-map}.a, Prologue and Epilogue are executed only once, while kernel is  executed multiple times instead.

The length of the kernel is called Iteration Interval (II) and lower II corresponds to lower execution times. Minimizing the II is the objective of every mapping algorithm, since lower II corresponds to high-quality mappings. Figure \ref{fig:cgra-map}.c shows how the DFG in Figure \ref{fig:cgra-map}.b can be scheduled and mapped on a $2\times 2$ CGRA with an II of 3. The minimum iteration interval (mII) is computed according to the formula in \cite{rau1996iterative}, that is:
\begin{equation}
mII = max(ResII, RecII)
\label{eq:mii}
\end{equation}
The first lower bound is given by the resource available on the CGRA and by the resource needed by the DFG, $ResII = \big\lceil \frac{\#nodes_dfg}{\# PE}\big\rceil = \big\lceil \frac{11}{4}\big\rceil = 3$. The second bound is given by the length of the longest loop $l$ in the DFG, $RecII = max\left(\lceil\frac{length(l)}{distance(l)}\rceil\right)_{l \in DFG} = 2$.

According to equation \ref{eq:mii} the lowest possible II for the DFG in Figure \ref{fig:cgra-map} is 3, so there is no possible better solution than the one shown in Figure \ref{fig:kms-map}.c.

\section{Methodology}

\subsection{Compilation}
\vskip -1em
\begin{figure}
    \centering
    \includegraphics[width=\textwidth]{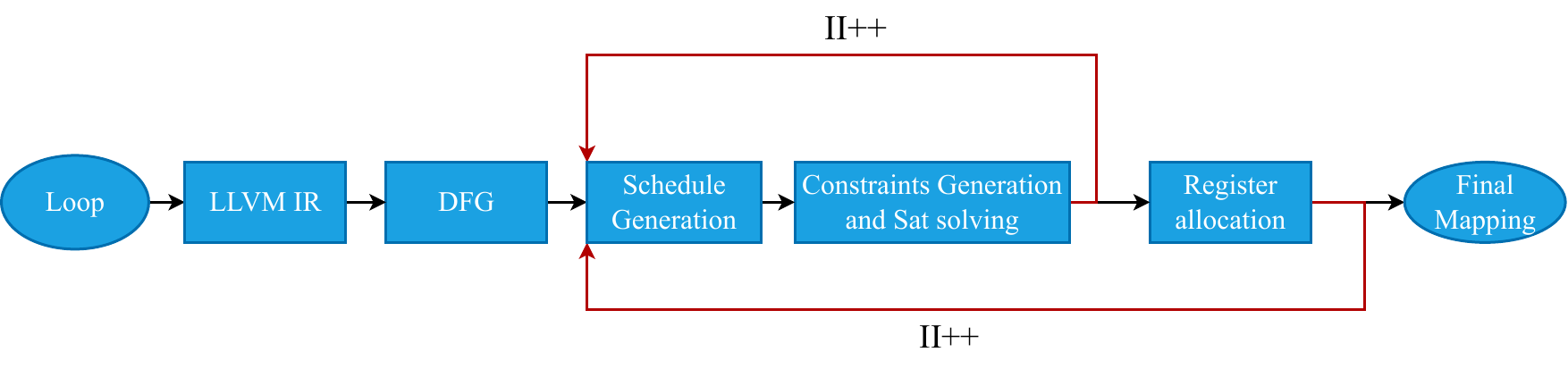}
    \caption{SAT-MapIt tool-chain iteratively increases the II if no mapping is returned by the SAT solver or if the register allocation phase failed}
    \label{fig:toolchain}
\end{figure}
The compilation process targets compute-intensive loops, which can be identified automatically or manually via pragma annotations. Since our research is focused more on the mapping phase, we manually mark the loops to be mapped with a pragma. 

As shown in Figure \ref{fig:toolchain} the first step of the compilation process we developed is to generate a semantically equivalent version of the loop with LLVM Intermediate Representation (LLVM IR) and from there extract the Data Flow Graph (DFG).

\begin{figure}
    \centering
    \includegraphics[width=\textwidth]{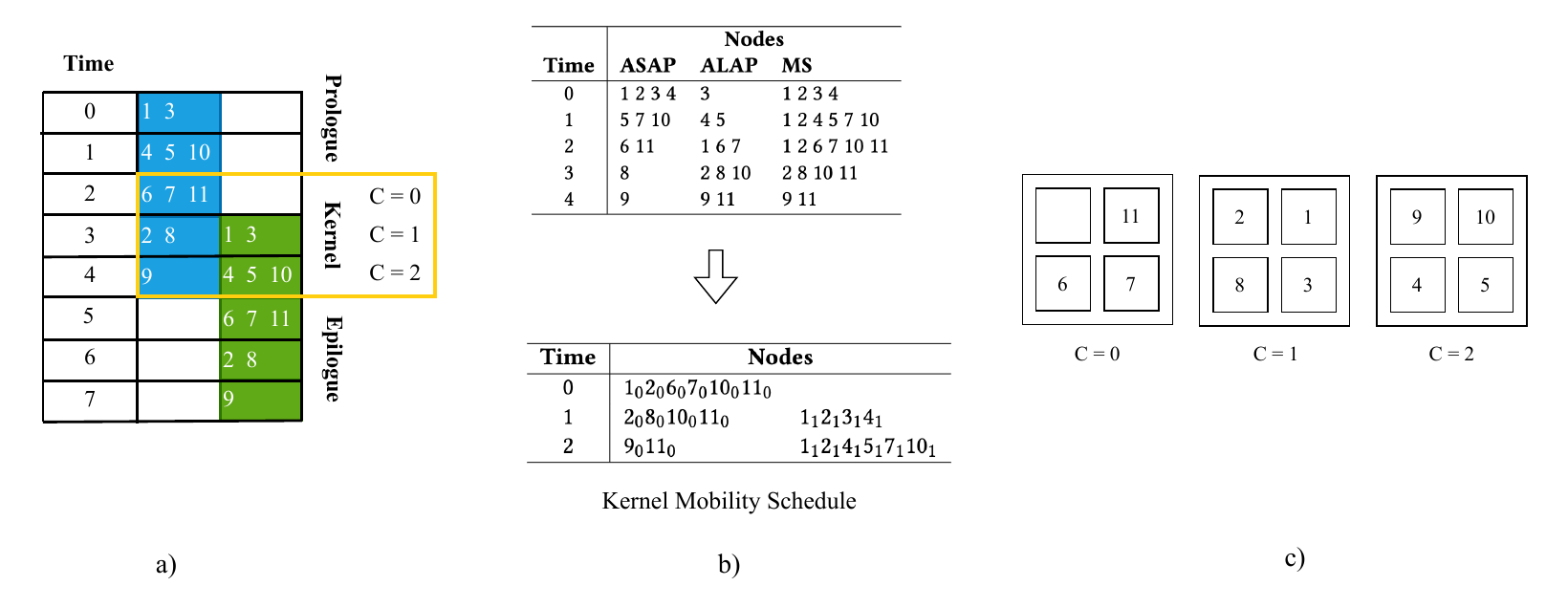}
    \caption{a) Modulo scheduling of the DFG in Figure \ref{fig:cgra-map}.b, highlighting the division between prologue, kernel, and epilogue. b) Kernel Mobility Schedule generation. c) One valid mapping of the kernel on a $2\times 2$ CGRA.}
    \label{fig:kms-map}
\end{figure}

Next, in the Schedule Generation phase we introduce a novel schedule called Kernel Mobility Schedule (KMS), which is necessary to our SAT formulation to find the mapping with the lowest II. The first step to generate the KMS is the construction of the Mobility Schedule (MS).
By using the As-Soon-As-Possible (ASAP) and As-Late-As-Possible (ALAP) schedules we create the MS, and then with a starting value for the II we iteratively fold the MS on itself by adding a label to every node after every folding. Figure \ref{fig:kms-map}.b  shows this step applied to the graph in Figure \ref{fig:cgra-map}.b.
The KMS is then used with the DFG to generate the constraints the SAT solver needs to find a valid solution.
Our formulation is register agnostic, so we need to have an additional phase, that is Register Allocation, to validate the output of the SAT solver. 
\vskip -1em
\subsection{Formulation}
Our mapping methodology leverages the SAT solver's ability to efficiently navigate the problem's complex solution space.
We encode the problem as conjunctive normal form (CNF) formula using literals in the form:
$x_{n, p, c, it}$ , where $n$ denotes the node identifier in the DFG, $p$ denotes a PE on the CGRA, $c$ represents at which cycle a node is scheduled, and $it$ to  which iteration the node refers to. 
From an high-level point of view all statements can be divided in three main sets of clauses:
\begin{itemize}
\item C1: Each node is associated with a set of literals. Within each set, exactly one literal must be assigned as \textit{True}. 
\item C2: Since a PE can only execute one instruction per cycle, at most one node can be mapped on a given PE at a given cycle.
\item C3: Every node's predecessor and/or successor must be placed in a neighbour PE, to ensure that data can be shared through the PE network.
\end{itemize}

This formulation can also use modulo scheduling to reduce the execution time on the CGRA thanks to the Kernel Mobility Schedule. The full and  more formal definition of the constraints is explained in detail in our recent work \cite{sat_extended,Tirelli2023}. 
\section{Experimental Results}
\vskip -1em
\begin{figure}[t]
    \centering
    \includegraphics[width=\textwidth]{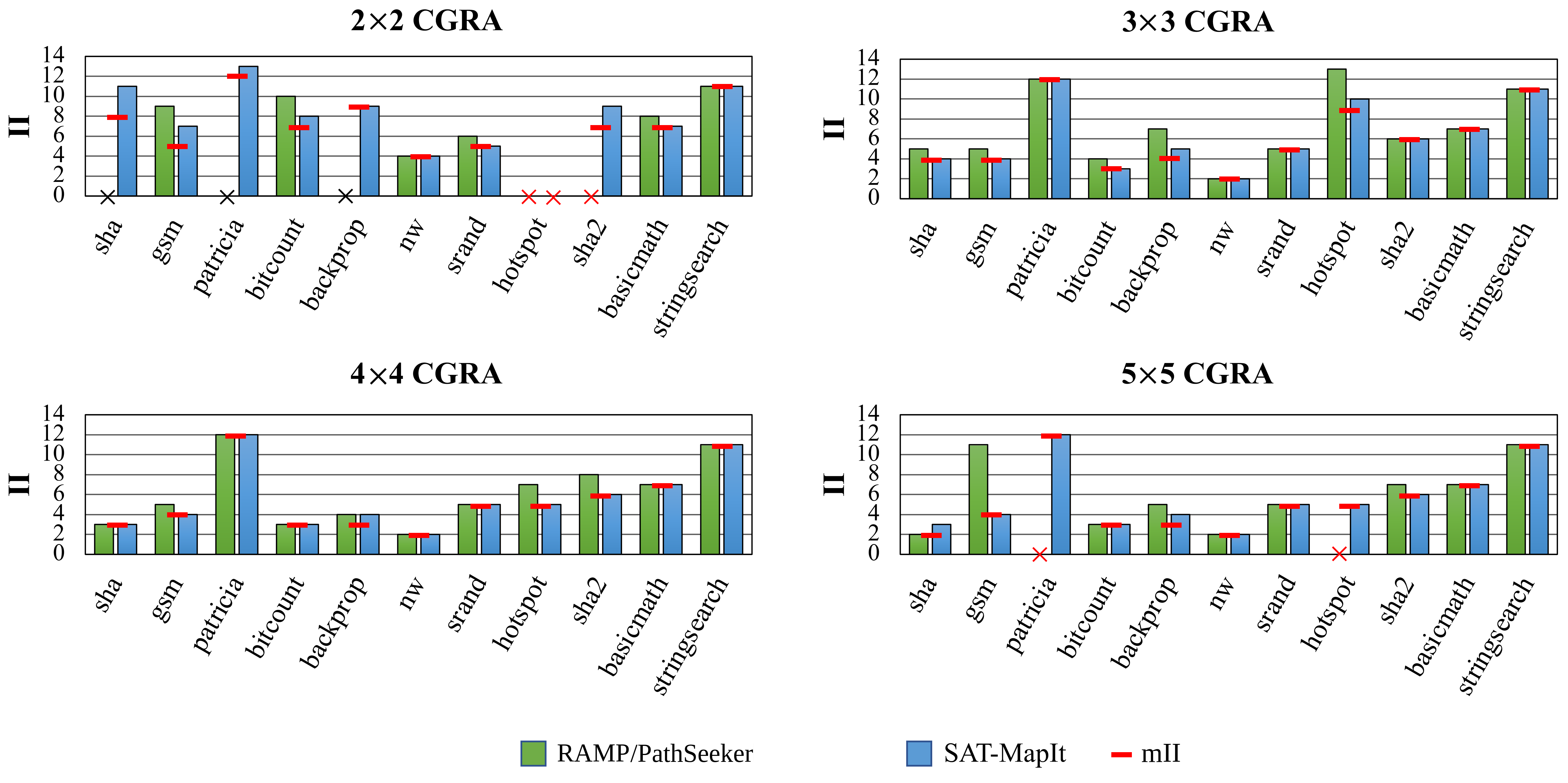}
    \caption{Experimental results of the chosen benchmarks for 4 different CGRA  sizes.
We compare the best II obtained by RAMP and PathSeeker with the II found by our tool-chain SAT-MapIt. A red cross means that the process did not terminate before a timeout of 4000 seconds. 
A black cross means the process was terminated when it reached the max II allowed (50) without finding a feasible solution.
Red dashes indicate the minimum II (mII). For the $2\times2$ CGRA, \texttt{hotspot} had a mII of 17, which is not displayed.}
    \label{fig:ii}
\end{figure}
Using a set of benchmarks from MiBench and Rodinia, we compare our tool-chain SAT-MapIt \cite{Tirelli2023} against the SoA tools RAMP\cite{dave2018ramp} and PathSeeker\cite{pathseeker}. Figure \ref{fig:ii} shows the II found by every tool for different CGRA sizes. Our methodology can explore the solution space fully and, consequentially, always return the best possible solution for every DFG considered.

It can be seen that SAT-MapIt outperforms the SoA in almost 50\% of the benchmarks by finding mappings with low II. Compilation times are also lower on average in our experiments.
In 26 out of 44 cases, our compilation time is is higher than the SoA, although the average time increase is only 15 seconds. While in the other cases, when our compilation times are lower, the time saved is on average 962 seconds.


\section{Conclusion}
During our research, we have introduced several novelties into the SoA. We proposed a new schedule called KMS and a novel SAT formulation that can exploit the modulo scheduling optimization technique to always return the best possible mapping for a given DFG. We also made open source the tool-chain (SAT-MapIt) \cite{satopenCT} that implements all the steps needed to map on a CGRA.
Our mapper's output is currently compatible with the \textit{OpenEdgeCGRA} \cite{Carpentieri2024}, but it can be easily adapted to any other architecture.

\subsection{Future work}
The focus of our research is centered on compilation technique for CGRA and we are currently trying to further expand our methodology to support larger DFGs and CGRA sizes. 
However, we are also exploring promising emerging technologies, such as neuromorphic computing architectures, which currently require significant enhancements in compilation techniques.

\subsubsection{\ackname} This study is  supported by the Swiss National Science Foundation under Grants: ML-Edge (200020-182009) and ADApprox (200020-188613).

%
%
\bibliographystyle{splncs04}
\bibliography{main}

@string{DAC17   = "Proceedings of the 54th Design Automation Conference"}

@string{DAC18   = "Proceedings of the 55th Design Automation Conference"}

@string{DATE    = "Proceedings of the Design, Automation and Test in Europe Conference and Exhibition"}

@string{IEEETCAD = "{IEEE} Transactions on Computer-Aided Design of Integrated Circuits and Systems"}

@string{TRETS= "ACM Transactions on Reconfigurable Technology and Systems (TRETS)"}

@string{ACM   = "ACM Sigplan Notices"}

@string{ACMOP = "Proceedings of the 20th ACM International Conference on Computing Frontiers"}

@string{IJPP   = "International Journal of Parallel Programming"}

@article{chen2014graph,
  title={Graph minor approach for application mapping on {CGRA}s},
  author={Chen, Liang and Mitra, Tulika},
  journal=TRETS,
  volume={7},
  number={3},
  pages={1--25},
  year={2014},
  publisher={ACM New York, NY, USA}
}

@article{pathseeker,
  title={Path{S}eeker: A {F}ast {M}apping {A}lgorithm for {CGRA}s},
  author={Balasubramanian, Mahesh and Shrivastava, Aviral},
  journal=DATE,
  year={2022},
  publisher={IEEE}
}

@inproceedings{dave2018ramp,
  title={R{AMP}: {R}esource-{A}ware {M}apping for {CGRA}s},
  author={Dave, Shail and Balasubramanian, Mahesh and Shrivastava, Aviral},
  booktitle=DAC18,
  pages={1--6},
  year={2018},
  organization={IEEE}
}

@inproceedings{chin2018architecture,
  title={An {A}rchitecture-{A}gnostic {I}nteger {L}inear {P}rogramming {A}pproach to {CGRA} {M}apping},
  author={Chin, S Alexander and Anderson, Jason H},
  booktitle=DAC18,
  pages={1--6},
  year={2018}
}

@article{satopenCT,
  title={SAT-{M}ap{I}t: {A}n {O}pen {S}ource {M}odulo {S}cheduling {M}apper for {C}oarse {G}rain {R}econfigurable {A}rchitectures},
  author={Tirelli, Cristian and Ferretti, Lorenzo and Pozzi, Laura},
  journal=ACMOP,
  pages={383–-384},
  year={2023},
  publisher={ACM}
}

@article{rau1996iterative,
  title={Iterative {M}odulo {S}cheduling},
  author={Rau, B Ramakrishna},
  journal=IJPP,
  volume={24},
  number={1},
  pages={3--64},
  year={1996},
  publisher={Springer}
}

@inproceedings{karunaratne2017hycube,
  title={Hy{CUBE}: {A CGRA} with reconfigurable single-cycle multi-hop interconnect},
  author={Karunaratne, Manupa and Mohite, Aditi Kulkarni and Mitra, Tulika and Peh, Li-Shiuan},
  booktitle=DAC17,
  pages={1--6},
  year={2017}
}

@article{li2021chordmap,
  title={Chord{M}ap: {A}utomated {M}apping of {S}treaming {A}pplications onto {CGRA}},
  author={Li, Zhaoying and Wijerathne, Dhananjaya and Chen, Xianzhang and Pathania, Anuj and Mitra, Tulika},
  journal=IEEETCAD,
  year={2021},
  publisher={IEEE}
}

@INPROCEEDINGS{Tirelli2023,
  author={Tirelli, Cristian and Ferretti, Lorenzo and Pozzi, Laura},
  booktitle={2023 Design, Automation \& Test in Europe Conference \& Exhibition (DATE)}, 
  title={{SAT-MapIt: A SAT-based Modulo Scheduling Mapper for Coarse Grain Reconfigurable Architectures}}, 
  year={2023},
  volume={},
  number={},
  pages={1-6},
  doi={10.23919/DATE56975.2023.10137123}}

@misc{openEdgeCGRA,
  author = {Beno\^it Denkinger et al.},
  title = {{ESL-CGRA Gitlab repository}},
  howpublished = "https://github.com/esl-epfl/OpenEdgeCGRA",
  year = {2020}, 
}

@inproceedings{Carpentieri2024,
 title={Performance evaluation of acceleration of convolutional layers on OpenEdgeCGRA},
  author={Carpentieri, Nicolo and Sapriza, Juan and Schiavone, Davide and Jahier Pagliari, Daniele and Atienza, David and Martina, Maurizio and Burrello, Alessio},
  booktitle={Workshop on Open-Source Hardware},
  pages={1--4},
  year={2024},
  organization={ACM}
}

@inproceedings{sat_extended,
 title={{SAT-based Exact Modulo Scheduling Mapping for Resource-Constrained CGRAs}},
  author={Tirelli, Cristian and et al.},
  booktitle={Journal on Emerging Technologies in Computing Systems},
  pages={1--26},
  year={2024},
  organization={ACM}
}

\end{document}